\documentclass[amsmath,amssymb,aip,
 reprint,twocolumn, superscriptaddress,twocolumn,
 amsmath,amssymb,
 aps  
]{revtex4-2}

\usepackage{graphicx}
\usepackage{dcolumn}
\usepackage{bm}
\usepackage[utf8]{inputenc}
\usepackage[T1]{fontenc}
\usepackage{mathptmx}
  
\usepackage[normalem]{ulem} 	
\usepackage[usenames,dvipsnames]{color}
\usepackage{xcolor} 
\usepackage{SIunits} 

\newif\iffinal
\finaltrue

\newcommand{\ket}[1]{\ensuremath{|{#1}\rangle}} 

\newcommand{\fres}[0]{\ensuremath{f_\mathrm{res}}} 
\newcommand{\fdrive}[0]{\ensuremath{f_\mathrm{d}}} 
\newcommand{\fqubit}[0]{\ensuremath{f_\mathrm{q}}} 
\newcommand{\ftls}[0]{\ensuremath{f_\mathrm{TLS}}} 
\newcommand{\geff}[0]{\ensuremath{g_\mathrm{eff}}} 
\newcommand{\ctot}[0]{\ensuremath{C_\mathrm{tot}}}
\newcommand{\phiq}[0]{\ensuremath{\Phi_\mathrm{q}}}

\newcommand{\del}[1]{\sloppy{\textcolor{red}{\sout{#1}}}} 

\newcommand{\new}[1]{\textcolor{red}{#1}}

\iffinal

\renewcommand{\del}[1]{}

\renewcommand{\new}[1]{#1}
\else
\fi

\makeatletter
\let\@fnsymbol\@fnsymbol@latex
\@booleanfalse\altaffilletter@sw
\makeatother

\begin{document}

\title{Readout failures in superconducting qubits due to TLS-defects in tunnel junctions}
\author{J\"urgen Lisenfeld}
\email[corresponding author, eMail: ]{juergen.lisenfeld@kit.edu}
\affiliation{Physikalisches Institut, Karlsruhe Institute of Technology, 76131 Karlsruhe, Germany}
\author{Alexander K. H\"andel}
\affiliation{Physikalisches Institut, Karlsruhe Institute of Technology, 76131 Karlsruhe, Germany}
\author{Alexander Bilmes}
\affiliation{Physikalisches Institut, Karlsruhe Institute of Technology, 76131 Karlsruhe, Germany}
\author{Alexey V. Ustinov}
\affiliation{Physikalisches Institut, Karlsruhe Institute of Technology, 76131 Karlsruhe, Germany}

\date{\today}

\begin{abstract}
	\centering\begin{minipage}{\linewidth}
		\textbf{
			Material defects give rise to parasitic two-level systems (TLS) which present a major source of decoherence in superconducting qubits. Here, we study a strongly coupled TLS that resides in the tunnel barrier of transmon qubit. 			
			We use multi-photon spectroscopy and TLS strain tuning to explore the rich spectrum of the interacting three-partite system consisting of TLS, qubit, and its readout resonator.			
			This reveals a strong effective resonant coupling between the TLS and the qubit's readout resonator which dresses the resonator states and results in a resonance frequency shift which spoils the readout signal. Our finding presents yet another way how material defects can interfere with qubit operation and hinder the realization of solid-state quantum processors.
			}
	\end{minipage}
\end{abstract}

\maketitle 
\setlength{\parskip}{-0.25cm}
\section{INTRODUCTION}  \vspace{-0.1cm}
Superconducting microcircuits are the most promising platform to realize solid-state quantum computers. However, their progress towards large and universally useful processors is severely hampered by decoherence~\cite{rao2026coherence}, of which the main part originates in elusive material defects that form parasitic two-level tunneling systems (TLS)\cite{Martinis2005}. Such defects occur e.g. in amorphous oxides and at disordered interfaces, and possess an electric dipole moment by which they couple to oscillating fields in qubits and resonators~\cite{muller2019towards}. TLS that are resonant with the qubit enhance energy relaxation, and their fluctuating resonance frequencies cause qubit instability and dephasing\cite{burnett2019decoherence,abdurakhimov2022identification}. Loss also occurs when driven qubits are brought into resonance with a TLS by the AC-stark shift\cite{carroll2022dynamics}, and via Landau-Zener transitions when the qubit is is tuned through TLS resonances.\\
\indent The detrimental impact of a TLS increases with its coupling strength to the qubit. TLS which are residing in the tunnel barrier of a qubit's Josephson junctions are most strongly coupled and give rise to avoided level crossings which can render the qubit unusable. Such qubit dropouts present a significant challenge to realize error correction in large-scale quantum processors~\cite{mohseni2024build}. In driven qubits or resonators, energy can leak even to far off-resonance TLS via unwanted multi-photon transitions~\cite{sank2016measurement,dai2026characterization}.\\

Here, we identify another source of readout error which occurs when a strongly coupled TLS in a qubit junction is in resonance with the readout resonator. This gives rise to a qubit-mediated effective TLS-resonator interaction \del{which dresses the resonator states and results in} \new{that creates an avoided-level-crossing in the resonator spectrum and causes}  resonance frequency shifts that spoil the readout signal. The effective coupling between the TLS and the resonator occurs via virtual qubit excitations. 
This is similar \del{to the direct coupling of TLS to drive signals
	, and} to the recent observation of a TLS in a coupler element of a quantum processor which interacted with two spatially distant qubits\cite{feiyan}. \del{In quantum processors, this so-called "cross-resonance" interaction is utilized to realize two-qubit gates.} 
\\

In our experiment, we use spectroscopy to directly observe the avoided-level crossing in the resonator spectrum that emerges from its \del{cross-resonant  interaction with} \new{qubit-mediated virtual resonant coupling to} a TLS in the qubit. This is possible by tuning the TLS resonance frequency via mechanical strain that is applied to the qubit chip~\cite{grabovskij2012strain,Lisenfeld19}. Moreover, spectroscopy at higher drive powers reveals a rich landscape of multi-photon transitions in the spectrum of the qubit-resonator-TLS system. These data agree well with \del{quantum} \new{numerical} simulations, which however did not allow us to reproduce all observed features. \del{due to the large size of the Hilbert space and accordingly long simulation time.} \\
\indent Our results demonstrate that the qubit-mediated coupling between TLS and resonators can lead to failure or degradation of qubit readout. For driven qubits, we show that this mechanism gives rise to a multitude of unwanted transitions by which quantum information can leak to environmental states.

\begin{figure*}[t!]
\includegraphics[width=\textwidth]{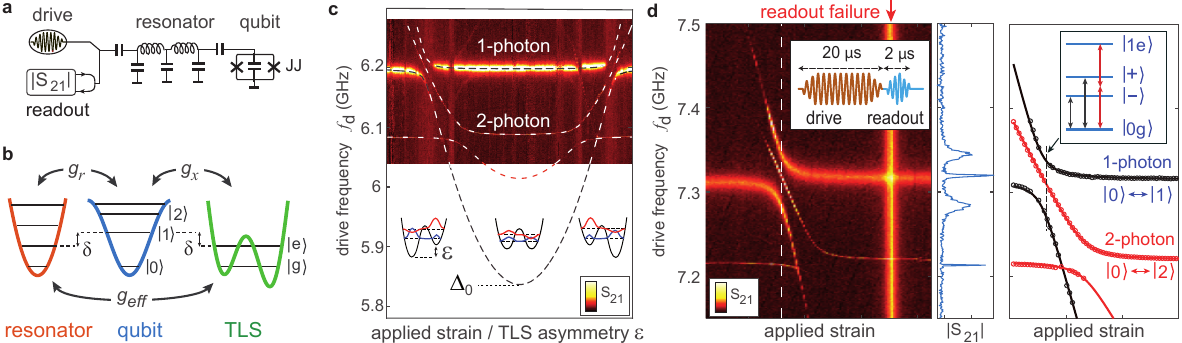}
	\caption{\textbf{a} Schematic of the transmon qubit circuit. A measurement of the transmitted signal $|S_{21}|$ at the resonator frequency is used for qubit readout.
		\textbf{b} Potentials of the resonator, qubit, and TLS, indicating their discrete eigenstates, mutual interactions $g_r$ and $g_x$, the qubit's detuning to resonator and TLS $\delta$, and the qubit-mediated effective resonator-TLS interaction $\geff$.
		\textbf{c} Pump-and-probe qubit spectroscopy observing both 1-photon $|0\rangle\rightarrow|1\rangle$ and 2-photon $|0\rangle\rightarrow|2\rangle$ transitions (horizontal traces). 	
		The color scale encodes the transmitted signal amplitude $|S_\mathrm{21}|$ at the readout resonator frequency.
		A sweep of the mechanical strain reveals avoided-level crossings in both transitions due to a strongly coupled TLS. The fit (dashed lines) provides the TLS' gap energy $\Delta$,  strain coupling strength, and its coupling electric dipole moment. The insets illustrate how the strain affects the asymmetry of the TLS' double-well potential.
		\textbf{d} Zoom into one anti-crossing (left panel), and the cross section along the white dashed line (middle panel). The red arrow indicates the parasitic readout signal which occurs when the TLS is tuned into resonance with the readout resonator.
		The right panel shows the peak positions (circles) and a fit to numerically calculated single-photon (black) and two-photon (red) transitions. The inset illustrates the energy level diagram and transitions of the qubit-TLS system at resonance.
	}
	\label{fig:1}
\end{figure*}

\section{METHOD} \vspace{-0.1cm}
We studied a tunable transmon qubit fabricated from aluminium on a silicon substrate, following the design of Barends et al.~\cite{Barends13}. The qubit is capacitively coupled to a readout resonator as sketched in the circuit schematic Fig.~\ref{fig:1}a to realize conventional dispersive readout \cite{Koch_2007}. The sample was fabricated as described in Refs.~\onlinecite{burnett2018noise,burnett2019decoherence} by D.~P. Lozano in the group of J. Bylander at Chalmers University, using a process that avoids unwanted stray junctions~\cite{osman2021simplified}.
 Table~\ref{table:fitparms} summarizes the qubit parameters, and more details on the setup are given in Appendix A.\\ 
\indent While TLS defects have long been known to limit qubit coherence, their microscopic origins are still unresolved. However, they can be described by a charge that can tunnel between the two minima of a generic double-well potential. The charge displacement creates an electric dipole moment by which TLS couple to the oscillating electric fields of the qubit mode. 
The TLS' potential is described by its tunneling energy $\Delta_0$ and its asymmetry energy $\varepsilon$ as illustrated in Fig.~\ref{fig:1}c.\\

Our setup allows us to tune TLS resonance frequencies via an applied mechanical strain that is generated by a piezo stack that exerts force on the backside of the qubit chip. The resulting mechanical strain $\mathcal{E}$ is of order $10^{-6}$ per applied Volt on the piezo~\cite{grabovskij2012strain}, and linearly modifies the asymmetry energy $\varepsilon = \gamma\cdot \mathcal{E}$ of the TLS' double-well potential at typical strengths~\cite{muller2019towards} of order $\gamma\approx 1$eV. This tunes the TLS resonance frequency along the hyperbola $f_\mathrm{TLS} = (\Delta_0^2 + \varepsilon^2)^{1/2}/h$, and allows one to fully characterize TLS properties by tracking their response with qubit energy relaxation spectroscopy. In the past, we used this method i.a. to study TLS densities in Josephson junctions\cite{Lisenfeld19} and near qubit electrodes\cite{bilmes2022probing}, to reveal mutual TLS interactions\cite{Lisenfeld2015}, and to characterize the decoherence of TLS due to interactions with phonons and thermally fluctuating TLS \cite{lisenfeld2016decoherence}.

\section{Results }\vspace{-0.2cm}
Here, we study a strongly coupled TLS with qubit spectroscopy by applying a microwave pulse of varying drive frequency \fdrive\ and fixed duration 20\,\textmu s that is much longer than the qubit $T_1$-time to reach a steady state. Any population of excited qubit states is then detected from the dispersive shift~\cite{Koch_2007} of the readout resonator that is proportional to the transmitted amplitude $|S_{21}|$ of a successive readout pulse at the resonator frequency \fres. In our measurements, the qubit is kept at a fixed frequency while the TLS is strain-tuned through the qubit resonance to observe avoided level crossings in their spectrum. To also observe multi-photon transitions to higher excited qubit states, the power of the qubit spectroscopy pulse is increased~\cite{wallraff2003multiphoton}. More details on experimental procedures are given in Appendix C.\\

Figure~\ref{fig:1}c shows avoided level crossings in both the qubit's single-photon $|0\rangle\rightarrow|1\rangle$-transition (here at 6.2 GHz), and two-photon $|0\rangle\rightarrow|2\rangle$ transition (here at 6.08 GHz). Such data are fitted with the help of the QuTiP software package~\cite{qutip1,qutip2} to obtain the TLS' parameters (see Table~\ref{table:fitparms}) and to identify the number of involved photons in the transition.\\

Figure~\ref{fig:1}d shows a zoom onto the left qubit-TLS anti-crossing when the qubit was tuned to 7.32 GHz. Notably, in such measurements we observed a strong readout signal appearing at a certain strain value independent of applied drive frequency (red arrow in Fig.~\ref{fig:1}d).
As detailed in the following, this is the signature of the readout failure that occurs when the TLS is tuned into resonance with the resonator where an effective strong resonant coupling between TLS and readout resonator is mediated via virtually excited qubit states. \new{This coupling results in an avoided level crossing in the resonator spectrum which superimposes the dispersive resonator shift due to its coupling to the qubit that serves as the qubit readout signal.}\\

\begin{figure*}[htbp]
	\centering
	\includegraphics[width=\textwidth]{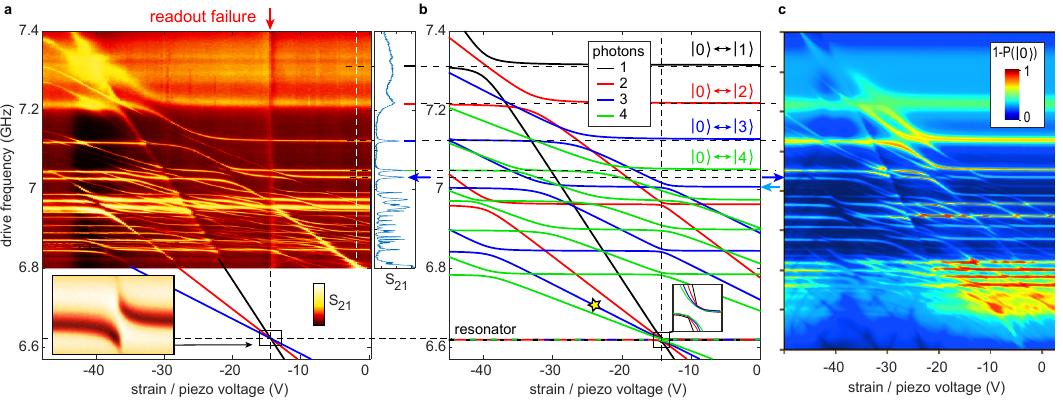}
	\caption{\textbf{a} Spectroscopy of the qubit-resonator-TLS system at enhanced drive power to observe multi-photon transitions involving higher excited states. The right inset shows the cross-section at the white dashed line. Due to the high drive power, the qubit's $|0\rangle \rightarrow |1\rangle$ resonant peak at 7.3 GHz is significantly power-broadened. The tilted transitions involve TLS states whose extrapolations intersect at the resonator frequency, illustrating that the readout failure (red arrow) occurs when the TLS is resonant with the resonator at 6.63 GHz. The left inset shows a direct measurement of the resonator anti-crossing without an applied drive tone.		
	\textbf{b} Transitions between eigenstates of the non-driven system as calculated with QuTiP. The line color indicates the number of involved photons. 
	\textbf{c} Simulation of the full time evolution of the system, including the driving pulse, energy relaxation, and six energy levels in both qubit and resonator. The color encodes the qubit excitation probability $(1 - P(\ket{0}))$ after reaching a steady state. Blue arrows indicate a closer fit to higher-order transitions compared to \textbf{b} (blue arrows).
}\label{fig2}
\end{figure*}

To obtain further insights into the interacting qubit-resonator-TLS system, we increase the spectroscopy drive power and extend the investigated frequency range. This reveals a rich spectrum of multi-photon transitions involving various higher-excited states as shown in Fig.~\ref{fig2}a. 
The origin of these transitions is clarified with a QuTiP calculation of the system's eigenenergies shown in Fig.~\ref{fig2}b. 
The line color indicates the number of photons involved in a transition. For example, from energy conservation we find that the red and horizontal two-photon trace at 6.955  GHz corresponds to the simultaneous excitation of the qubit (at 7.3 GHz) and the resonator (at 6.61 GHz).\\
\indent Transitions which involve the TLS appear tilted due to the TLS' strain tunability, while the tilt angle decreases with the photon number. For example, at the transition marked by a star in Fig.~\ref{fig2}b, three photons of energy $h\cdot6.7$~GHz are absorbed of which two photons end up in the resonator and one photon excites the TLS which has a resonance frequency of 6.88 GHz at the corresponding strain value.\\

When the tilted TLS transitions are extrapolated, they intersect at the resonator frequency where they cause an avoided-level crossing (see inset in Fig.~\ref{fig2}b). We directly confirm the emergence of the avoided-level crossing by strain-dependent resonator spectroscopy without driving the qubit (see inset of Fig.~\ref{fig2}a). The resulting shift of the resonator frequency is the origin of the parasitic readout signal (red arrows in Figs.~\ref{fig:1}d and ~\ref{fig2}a) and the cause of qubit readout failure.
The data in Fig. 2a thus simultaneously tracks the strain-dependent TLS frequency and confirms that the parasitic signal appears when the TLS is resonant with the resonator.\\

\begin{table} [b]
	\centering
	\begin{tabular}{llll}
		\hline
		name & value & description\\ 
		\hline 
		$f_\mathrm{q,max}$ & 7.32 GHz & maximum qubit frequency\\ 
		$E_{c}/h$  & 228 MHz & qubit charging energy\\ 		
		$A_\mathrm{JJ}$ & 200 nm x 220 nm & JJ area\\ 
		\hline
		$f_\mathrm{res}$ & $\approx$ 6.625 GHz & resonator frequency\\ 
		$g_\mathrm{qr}$ & 34 MHz & resonator-qubit coupling strength\\
		\hline
		$\gamma$ & 212 MHz / $V_\mathrm{piezo}$ & TLS strain-coupling strength \\ 
		$\Delta_0$ & $h\cdot 5.838$ GHz & TLS gap energy\\ 
		$\bar{p}$ & $0.36\,e$\AA\ & TLS dipole moment component\\
		$g_{\mathrm{x}}$ & 21.7 MHz & transversal coupling TLS-qubit\\
		$g_{\mathrm{z}}$ & <1 MHz (not measurable)& longitudinal coupling TLS-qubit\\
		\hline    
	\end{tabular}
	\caption{Parameters of the studied qubit circuit and the TLS.}
	\label{table:fitparms}
\end{table}

A closer comparison between the data in Fig.~\ref{fig2}a and the QuTiP calculation of eigenvalues shown in Fig.~\ref{fig2}b reveals discrepancies in higher-order transitions. For example, the 3-photon transition at $\approx$~7.03 GHz (marked by a blue arrow in Fig.~\ref{fig2}a) is measured at higher frequency than the eigenvalue solution predicts (light blue arrow in Fig.~\ref{fig2}b). 
To take into account possible AC-stark shifts due to the strong driving, we simulated the full time-evolution of the system including energy relaxation and the driving pulse by iteratively solving the Lindblad master equation with QuTiP. The resulting qubit expectation value is plotted in Fig.~\ref{fig2}\textbf{c} and achieves a close fit to the data (details on the simulation and measurements are given in Appendix B. Due to the \new{inefficient iterative master equation simulation in a large}\del{large size of the} Hilbert space spanned by 6 qubit states, 6 resonator states, and 2 TLS states, this simulation approach required a computation time of about 60 days on a 60-CPU cluster. However, various higher-order transitions are still only resolved in the experiment. \del{Since such measurements can be obtained within less than an hour, this might be seen as an example of the advantage of physical quantum simulations already for small multi-level systems.}\\

\begin{figure*}[htbp]
	\centering
	\includegraphics[width=\textwidth]{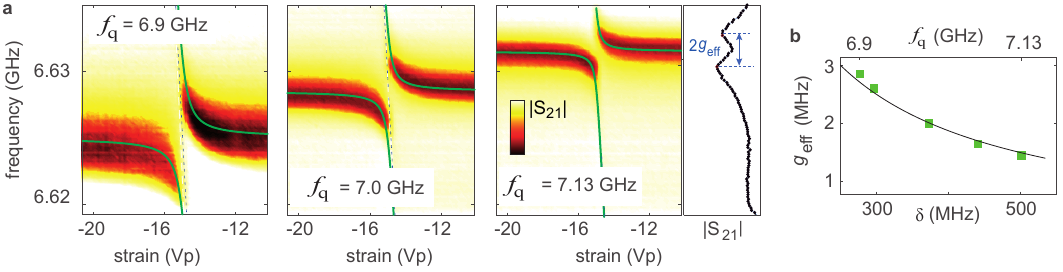}
	\caption{\textbf{a} Anti-crossings between the readout resonator and the strain-tuned TLS, measured when the qubit was tuned to the indicated resonance frequencies. The effective resonator-TLS coupling strength \geff\ is estimated from a fit (green line) and corresponds to half the distance between the two peaks at resonance as shown in the inset. \textbf{b} TLS-resonator coupling strength \geff\ vs. the detuning $\delta = \fqubit - \fres$ between qubit and resonator.  The solid line shows the expectation  from the \del{cross-resonance} \new{qubit-mediated} interaction, $g_\mathrm{eff} = g_x g_r / \delta$, for independently measured $g_r$ and $g_x$.
	}	\label{fig3}
\end{figure*}

A characteristic of the \del{cross-resonance} \new{coupling mediated by virtually excited qubit states  } is~\cite{feiyan} 
that the effective interaction strength between resonator and TLS is given by $\geff = g_r\cdot g_x/\delta$, where $g_r$ is the coupling strength of the resonator to the qubit,  $g_x$ is the transversal qubit-TLS coupling strength, and $\delta = \fqubit - \fres = \fqubit - \ftls$ is the detuning between qubit and resonator while $\delta \gg g_r,\, g_x$.
This is confirmed by spectroscopic measurements of the resonator-TLS anti-crossing when the (undriven) qubit is tuned to different resonance frequencies as shown in Fig.~\ref{fig3}\textbf{a}. Up to the largest investigated detuning $\delta=500$ MHz, the observed \geff\ exceeded 1.5 MHz. Figure~\ref{fig3}b shows an excellent agreement between the observed and expected $\geff$, where the latter is calculated using independently determined qubit-resonator and qubit-TLS coupling strengths (see Appendix C for details).\\

When the qubit-TLS coupling $g_x$ is comparable to the qubit-resonator coupling $g_r$ (typically,  $g_r\, \approx$ 50 - 100 MHz), the interaction with the TLS detunes the resonator by a similar amount as the dispersive shift $\chi = g_r^2/\delta$ 
that serves as the qubit readout signal\cite{Koch_2007}. Coupling strengths of $g_x\approx 10 - 40$ MHz are frequently observed for TLS in tunnel barriers of qubit junctions, where they occur at densities between 100 to 8000 TLS/(GHz$\cdot $\textmu m$^3$)\cite{Martinis2005,bilmes2022probing,osman2023mitigation,Wallraff2025,daum2025investigation,wolff2026structural} \new{in the 2-nm-thick tunnel barrier having typical areas of 200\,nm $\times$ 200\,nm}. In the studied sample, we have also detected a second TLS which caused similar readout failure (see Appendix C).\\

\new{\indent To probe for TLS-related readout degradation when mechanical strain-tuning is not available, conventional resonator spectroscopy (as in Fig.~\ref{fig3}) can be used to reveal a split resonance due to the virtually coupled TLS. As an alternative to applied mechanical strain, TLS can also be tuned by a DC-electric field~\cite{lisenfeld2019electric}. However, the E-field only tunes TLS that are located in vicinity of the qubit electrodes which typically interact much more weakly (factor $\approx$ 1/100) with the qubit than TLS in tunnel barriers. This reduces the effective TLS-resonator coupling $\geff$ accordingly and would not result in a clearly visible anti-crossing or similarly severe readout failures. Also, thermal cycling of the qubit sample redistributes the TLS spectrum~\cite{Lisenfeld2015}, which may serve as a test for a TLS-related origin of readout degradation.}\\

\indent The qubit-TLS coupling is usually modeled as a transverse (exchange) interaction where the TLS' dipole moment couples to the qubit charge. Alternatively, TLS might also modify the critical current of the junction~\cite{simmonds2004,abdurakhimov2022identification} and thus the qubit energy, corresponding to a longitudinal coupling. Since Cooper-pair tunneling is presumably dominated by a few channels which connect the thinnest parts of the tunnel barrier, the state of a charged TLS can affect the transparency of a channel and thus the critical current. The effect of such coupling is a small shift of the two-photon transition frequency in the resonantly coupled qubit-TLS system~\cite{ashhab2006rabi}. We have fitted our spectroscopy data including $g_z$ as a free parameter, which did not result in any detectable longitudinal coupling for the studied TLS at an estimated detection threshold of about $g_z \gtrapprox$ 3 MHz (see App. B and C for details). This is similar to previous results obtained on flux-\cite{lupacscu2009one} and phase-qubits~\cite{cole2010quantitative,bushev2010multiphoton}.

\section{DISCUSSION AND CONCLUSION}\vspace{-0.2cm}
We have revealed a mechanism by which a TLS defect in the tunnel barrier of Josephson junctions can give rise to qubit readout failure.
Spectroscopy of strain-dependent transitions in the qubit-resonator-TLS system confirms that this problem emerges from an effective strong resonant TLS-resonator interaction that is mediated via virtual qubit states. Using multi-photon spectroscopy, we map out the plethora of higher-order transitions that emerge from this interaction, and analyze them with the help of \del{quantum}\new{numerical} simulations.\\

\indent The parasitic TLS-resonator interaction results in a detuning of the resonator frequency which, for strongly-coupled TLS in qubit junctions, is of similar size as the dispersive shift that is used for qubit readout and thus spoils the measurement. In addition, the resonator \del{inherits} \new{may inherit} excess energy relaxation from its interaction with TLS that have typical quality factors below 5000~\cite{lisenfeld2016decoherence, Lisenfeld19}. Also, TLS resonance frequencies are unstable\cite{,klimov2018fluctuations,Wallraff2025} due to their interaction with neighboring thermal TLS and trapped charges\cite{faoro2014generalized,muller2015interacting,bilmes2017electronic}, and due to local strain changes caused e.g. by ionizing radiation impacts\cite{thorbeck2023}. This can give rise to excess resonator phase noise and temporal fluctuations of readout fidelity.\\
\indent These issues occur whenever a strongly-coupled junction-TLS is in resonance with the readout resonator. This has the same probability as the case that a qubit is inoperable due to strongly coupled junction-TLS, which occur at spectral densities of $\approx 0.1 - 2$ TLS per GHz for typically used small-area junctions~\cite{Wallraff2025,wolff2026structural}. For quantum processors, this may explain excessive readout errors observed in some qubits~\cite{morvan2024phase}.\\

\indent This problem is relevant to quantum processors and the large variety of qubit implementations which utilize dispersive readout. It can also play a detrimental role in bosonic or cat-qubits~\cite{grimm2020catQubit} where quantum information is encoded in superpositions of resonator states. 
Our results thus show yet another mechanism by which TLS hamper the functionality of superconducting qubits, and further motivate systematic studies to understand how defects can be avoided in qubit fabrication.\\ 

\noindent \textbf{Acknowledgements.} We thank J. Bylander and Daniel P. Lozano for providing the qubit sample. We acknowledge the computing time provided on the high-performance computer HoreKa by the National High-Performance Computing Center at Karlsruhe Institute of Technology, which is jointly supported by the Federal Ministry of Education and Research and the Ministry of Science, Research and the Arts of Baden-W\"urttemberg, as part of the National High-Performance Computing (NHR) joint funding program. HoreKa is partly funded by the German Research Foundation (DFG).\\ \indent This work was funded by Google, which we gratefully acknowledge.\\

\noindent \textbf{Author contribution.} JL conceived this work and performed the measurements. Data analysis and QuTiP simulations were done by KH and JL. AB took part in the development of the experimental setup and sample preparation. The experimental infrastructure was provided by A.V.U. The manuscript was written by JL with contributions from all authors.\\

\noindent \textbf{Data availability.} The data acquired during this work are available from the corresponding author upon reasonable request.\\

\onecolumngrid
\normalsize
\renewcommand{\thesection}{APPENDIX \Alph{section}}
\renewcommand{\thetable}{A\arabic{table}}
\renewcommand{\figurename}{Fig.}
\setcounter{section}{0}


\section{Experimental setup}
\phantom{hi there}\\

\label{app:setup}
\noindent
\onecolumngrid
\noindent\begin{minipage}[b]{0.3\textwidth}
	\vspace{0pt} 
	Figure 4 illustrates the experimental setup with standard components for dispersive qubit readout and qubit flux biasing. Mechanical strain is applied via a piezo stack actuator, which pushes on the chip backside. The resulting mechanical strain is of order $10^{-6}$ per applied Volt on the piezo~\cite{grabovskij2012strain}.\\		
	
	The qubit sample was fabricated from Aluminium on Silicon in the group of J. Bylander at Chalmers University. As shown in Fig. 5, the qubit consists of a cross-shaped island~\cite{Barends13} that is connected by two Josephson junctions to the surrounding ground plane. The Manhattan-style junctions have areas of about 200nm x 220nm.  The qubit frequency is tuned via an on-chip flux line.\\	
	\\		
    All measurements were done at a sample temperature of 25 - 30 mK.\\
\end{minipage}
\begin{minipage}{0.7\textwidth}
	\vspace{-8cm}
	\hspace{0.6cm}
\includegraphics[width=0.95\linewidth]{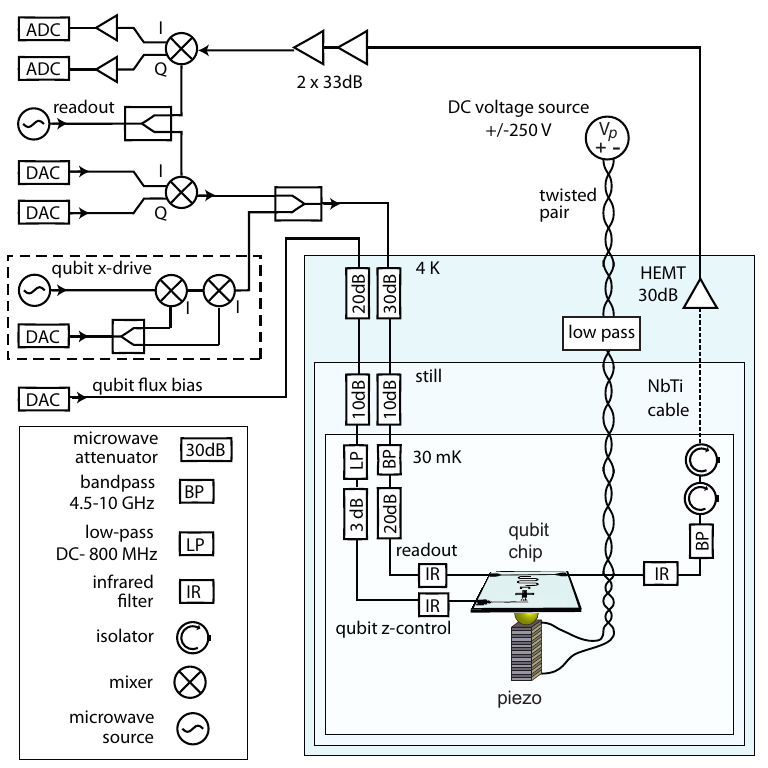}\\
	\phantom{hi there}\small Fig. 4. Schematic of the experimental setup. 
\end{minipage}
\phantom{hi there}\\
\setcounter{figure}{4}

\twocolumngrid
\phantom{hi there}\\
	\vspace{-1cm}
	\phantom{hi there}\\
	The energy relaxation time $T_1$ of the here investigated sample was very short ($T_1 \approx$ 1 - 2\textmu s), while other qubits on the same chip showed $T_1$-times of up to 28 \textmu s.  This is explained by the strongly coupled junction-TLS, whose detuning to the qubit is $\delta = |(\ftls - \fqubit| \lesssim 1$ GHz in a wide part of the investigated strain-tuning range. An estimation for the qubit $T_1$-time  due to the coupling to this TLS is~\cite{Barends13}
	\begin{equation*}
		T_1^{-1} = \frac{4 \pi g_x \Gamma}{\Gamma^2 + \delta^2} + \Gamma_{1,q}
	\end{equation*}
	where $\Gamma_{1,q}^{-1}\approx 25\,$\textmu s is the assumed qubit $T_1$ time if it would not be coupled to the TLS (as observed in the other qubits on the chip), and $\Gamma= \Gamma_{1,\mathrm{TLS}}/2 + \Gamma_{2,\mathrm{TLS}} + \Gamma_{1,\mathrm{q}}/2 + \Gamma_{2,\mathrm{q}}$ is the sum of qubit and TLS relaxation and dephasing rates. Using typical TLS values~\cite{Barends13,Shalibo2010,Lisenfeld19,Lisenfeld2015,Bilmes_2021,chen2024} of $\Gamma_{1,\mathrm{TLS}} \approx \Gamma_{2,\mathrm{TLS}} \approx 1/100$\,ns, the measured qubit-TLS coupling strength $g_x\approx 22$\,MHz, and detunings of $\delta = 1 - 2$~GHz, results in qubit $T_1$-times between 1.0 and 3.7 \textmu s which agrees to the observed values.
	\begin{figure}[h!]
		\includegraphics[width=\linewidth]{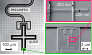}
		\caption{Scanning-electron microscopy of the qubit sample. The cross-shaped qubit capacitor is connected to the surrounding ground plane by a DC-SQUID made with two Manhattan-style tunnel junctions (insets).}
	\end{figure}

\section{Models} 
\label{app:simus}
\noindent\textbf{Qubit.} 
We follow Koch et al.~\cite{Koch_2007} and start from the transmon qubit Hamiltonian
$$H_{\mathrm{Tr}} =  4E_C\,\hat{n} - E_J\,\cos(\hat{\phi})$$ with the qubit charge and phase operators $\hat{n}$ and $\hat{\phi}$. The cosine is approximated for the transmon regime $\phi\ll 1$, and the qubit is modeled as an anharmonic (Duffing) oscillator
by dropping higher order and fast rotating terms: 
$$H_{Q} =  (\omega_Q + \alpha_q/2)\,q^{\dagger}q + \alpha_q/2\cdot(q^{\dagger}q)^{2},$$
where $\omega_Q = 2\pi\cdot \fqubit$ is the qubit resonance frequency and $\alpha_q$ is its anharmonicity. The qubit transition frequency $\fqubit$ depends on the flux $\phiq$ threading its DC-SQUID and can be approximated by\cite{Koch_2007}
\begin{equation*}
	h\fqubit(\phiq)=\sqrt{8E_C E_J(\phiq)}-E_C,
	\label{eq:qubitfreq}
\end{equation*}
where $E_C = e^2 / 2 C_\mathrm{tot} = -\alpha_q$ is the qubit's charging energy, $C_\mathrm{tot}$ is the combined junction and qubit island capacitance, $e$ is the electron charge, and
\begin{equation*}
E_J = \frac{\hbar I_c}{2e} \sqrt{\cos(\pi \phiq\Phi_0)^2+d^2\sin(\pi \phiq/\Phi_0)^2}
\end{equation*}
 is the flux-dependent Josephson energy assuming two slightly different junctions with a total critical current $I_c$, $\Phi_0=h/2e$ is the flux quantum, $h$ is Planck's constant, and the asymmetry of the junction's Josephson energies is described by $d=\frac{E_{J1}-E_{J2}}{E_{J1}+E_{J2}}$.\\

\noindent The \textbf{readout resonator} is represented by a harmonic oscillator and its interaction with the qubit is described by a the exchange coupling\\
\begin{eqnarray*}
&H_{R} = & \omega_R \cdot r^{\dagger}r \\
&H_{\mathrm{int,r}} = & h g_\mathrm{qr} \cdot (q + q^{\dagger}) \cdot (r + r^{\dagger})\\
\end{eqnarray*}
where $\omega_R = 2\pi\cdot \fres$ is the resonator frequency and $g_\mathrm{qr}$ is the qubit-resonator coupling strength. \\

\noindent The \textbf{TLS} is implemented as a pure two-level system, using harmonic ladder operators and restricting the Hamiltonian to two dimensions:
$$H_{\mathrm{TLS}} = \omega_{\mathrm{TLS}} \cdot t^{\dagger}t.$$
The qubit-TLS coupling is usually described as a transversal (exchange) interaction where the TLS' dipole moment couples to the qubit charge:
$$H_{\mathrm{int, x}} = h g_x \cdot (q + q^{\dagger}) \cdot (t + t^{\dagger}),$$
Alternatively, TLS might also modify the critical current of the junction~\cite{simmonds2004} and thus the qubit energy. Cooper-pair tunneling is presumably dominated by a few channels which connect the thinnest parts of the tunnel barrier, so that the state of charged TLS can affect the transparency of a channel as illustrated in the right part of Fig.~6. In this case the TLS might also be formed by charge traps, e.g. creating an Andreev-level fluctuator~\cite{faoro2005models,deSousa2009,muller2019towards}. However, experimental insights on the occurrence and role of critical-current fluctuators in qubits have so far remained elusive.\\

\noindent Via the coupling to the qubit's critical current, the TLS may modify the qubit energy which would result in a longitudinal coupling 
$$H_{\mathrm{int, z}} = g_Z \cdot (q^{\dagger}q) \cdot  (t^{\dagger}t).$$\\
\begin{center}
	\includegraphics[width=0.8\linewidth]{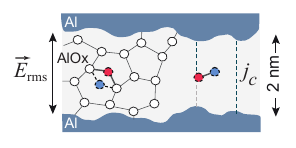}\\
\end{center}
\small{Fig. 6. Model of a Josephson junction with superconducting electrodes (blue) separated by a thin tunnel barrier (grey). The left part illustrates an atomic TLS which couples to the qubit charge via its electric dipole moment. The right part describes TLS coupling to the junction's critical current, where due to the electrode's roughness most of the supercurrent is carried by a few tunnel channels (dashed lines). A charged TLS may open or block a current channel and modify the junction's critical current.}\\[0.3cm]
\normalsize

\noindent \textbf{TLS-Qubit coupling.} The exchange coupling strength $g_x$ between qubit and TLS is given by the projection of the TLS' electric dipole moment onto the AC electric field from the qubit mode:
$$h g_x = \boldsymbol{pF} = \bar{p} |\boldsymbol{F}|,$$
 where $\boldsymbol{F}$ is the electric field inside the capacitor that is induced by the qubit plasma oscillation, and $\bar{p} = p_\parallel\, \Delta_0/(\hbar \omega_{\mathrm{TLS}})$ is the projection of the TLS' dipole moment $\boldsymbol{p}$ onto $\boldsymbol{F}$, multiplied by the TLS' matrix element $(\Delta_0/\hbar\omega_{\mathrm{TLS}}) $, i.e. the effective coupling.\\
For TLS in the tunnel barriers, $\bar{p}$ can be measured by determining $g_x$ from qubit-TLS swap oscillations or from anti-crossings in qubit spectroscopy, where the electric field strength $|F| = V_\mathrm{rms}/t$ in the junction is $V_\mathrm{rms}= \sqrt{\hbar \omega_{q} /2\ctot}$ at an estimated barrier thickness of $t\approx 2\,$nm.\\

\textbf{QuTiP simulations.}
We fit the acquired spectroscopy data with the help of the QuTiP package\cite{qutip1,qutip2}. For this, we first extract the positions of resonant peaks (see e.g. circles in Fig.~\ref{fig:1}d)). 
In a least-square fitting routine, the system's eigenvalues are calculated using the \texttt{qutip.eigenvalue()} function, and the energies of high-frequency transitions are divided by integer factors to enable fitting to the experimentally observed multi-photon transitions.
The model parameters are then optimized until the deviation between measured and simulated peak positions reached a minimum. 
This results in the solid lines shown in Fig.~1c,d, Fig. 2b, and Fig.~7. In the simulations, the  Hilbert space is restricted to 6 qubit states, 6 resonator states, and the 2 TLS states.\\

\noindent \textbf{Time-resolved simulation.} The comparison between the data in Fig. 2a and the eigenvalue fit shown in Fig. 2b reveals discrepancies in higher-order transitions. To take into account AC-stark shifts, we also simulate the full time evolution of the system. For this, the microwave spectroscopy is modeled by adding a drive term to the qubit, and QuTiP's \texttt{mesolve()}-function is used to iteratively solve the Lindblad master equation at consecutive time steps until Rabi oscillations subside and the system reached a stationary state. This also accounts for decoherence, for which corresponding collapse operators are applied whose amplitude is adjusted to the observed qubit and resonator quality factors.\\
\indent The time-resolved simulation indeed shows shifts in higher-order transitions that are similar to the observed data. However, due to the intense CPU-time requirements simulating interacting multi-level quantum systems, \new{and mostly because of our inefficient simulation method to find the steady-state of the system,} this approach failed to systematically fit the effects of the AC-stark shift to the data.\\

\section{Sample characterization}
\label{app:samplecharacterization}
The parameters of the system are listed in Table 1 of the main manuscript and are obtained as follows.\\

\textbf{Qubit readout.}
To detect the qubit level population, we measure the transmission $S_{21}$ at the resonance frequency of the readout resonator. For small qubit-resonator coupling strength $g_\mathrm{qr}$ and sufficiently large detuning $|\delta| = |\fqubit - \fres| \gg g_\mathrm{qr}$, an excitation in the qubit will change the resonator frequency by the dispersive shift\cite{Koch_2007}
$$	\chi\approx -\frac{ g_\mathrm{qr}^2}{\delta},$$
and this results in a detectable change of the transmission signal $S_{21}$.\\

\noindent\textbf{Qubit capacitance.} The total qubit capacitance is calculated from the anharmonicity observed by higher-power spectroscopy such as shown in Fig.~1d. We estimate it by fitting multi-photon qubit spectroscopy data to QuTiP simulations as shown in Fig.~1d, which results in $\ctot = e^2/2 E_c$ = 85 fF for Ec = 228 MHz.\\

\noindent\textbf{Qubit-resonator coupling strength}
We sweep the applied qubit flux and readout frequency to observe anti-crossings in the resonator spectrum (see Fig. 7). The data is then fitted with QuTiP to the model excluding the TLS with the free parameters $\fqubit,\, \fres,\, g_{q,r}$, junction asymmetry $d$, and a flux-scaling factor. This results in a qubit-resonator coupling strength $g_{q,r}\approx 34$~MHz and a large junction asymmetry of $d\approx 0.67$.\\

\noindent\textbf{Qubit-TLS coupling strength}
The exchange coupling strength $g_x$ between qubit and TLS is measured by sweeping the TLS through the qubit resonance via the applied mechanical strain, and fitting the anti-crossing to a QuTiP simulation as shown e.g. in Fig.~1d. The free parameters of the fit were $\fqubit,\, \alpha_q,\, \ftls,\, \fres,\, g_x,\, \gamma$ and $V_0$, where $V_0$ is the piezo voltage corresponding to the TLS' symmetry point $\varepsilon=0$. Fits to different data sets acquired at several drive powers (e.g. data shown in Fig.~1c,d, Fig.~2a) and qubit resonance frequencies (see Fig. 3) provide consistent values of $g_x \approx 22$\,MHz.
This corresponds to a parallel TLS dipole moment component of 
$\bar{p} = h\,g_x / |\boldsymbol{F}| \approx 0.36\,e$\AA\ with the electric field strength in the tunnel barrier $|\boldsymbol{F}| = \sqrt{\hbar \omega_{q} /2\ctot}/t$ and $t\approx 2\,$nm.\\

{\centering \includegraphics[width=0.8\linewidth]{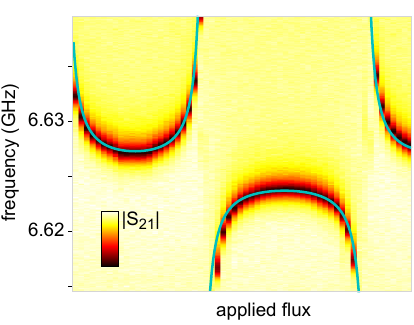} \par}

\small{\vspace{0.2cm} Fig. 7. Resonator spectroscopy, showing its anti-crossing with the flux-tuned qubit. The fit (cyan line) provides a qubit-resonator coupling strength of $g_r$=34 MHz and the resonator frequency $\fres = 6.6255$\,GHz.\\
\normalsize
\vspace{0.2cm}

\noindent \textbf{Critical-current coupling.} To check whether the TLS couples to the juntion's critical current, we repeated above fits and included the longitudinal TLS-qubit coupling strength $g_y$ as a free fit parameter. The effect of such coupling is a small shift of the two-photon transition frequency in the resonantly coupled qubit-TLS system. In similar previous analyses using flux- and phase-qubits\cite{lupacscu2009one,cole2010quantitative,bushev2010multiphoton}, no evidence for nonzero $g_y$ was found. In the here-studied transmon qubit, this effect is presumably amplified due to its smaller junction size. However, the fits resulted in values $g_y \lesssim 1$\,MHz that are smaller than the fit uncertainty, so that we found no evidence for a longitudinal coupling in the investigated TLS.\\
\indent Fig.~8 shows a QuTiP calculation of the qubit-TLS anti-crossing for various values of their longitudinal coupling strength $g_z$. The detuning of the two-photon line in the center of the anti-crossing scales as $\delta f = g_z/2$ (see inset of Fig.~8). The minimum detectable $g_z$ in this experiment can be estimated from the full width at half maximum
of the two-photon transition as shown in Fig.~1d, which is about $\approx 3$~MHz.\\

\begin{center}
	\includegraphics[width=0.8\linewidth]{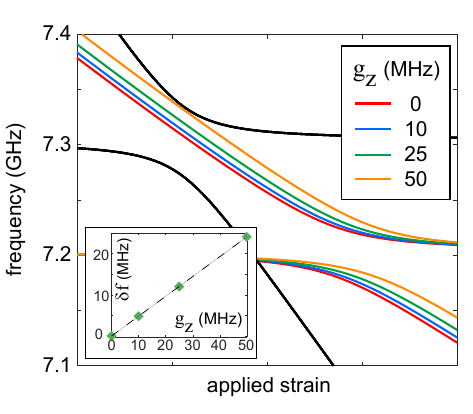}\\
\end{center}
\small{Fig. 8. Shift of the two-photon transition (colored lines) through the main qubit-TLS anti-crossing (black lines)  for various longitudinal TLS-qubit coupling strengths $g_z$. Inset: simulated frequency shift $\delta f$ at qubit-TLS resonance vs. $g_z$ (points), and a plot of $\delta f = 0.5\cdot g_z$ (dashed line).}\\[1cm]

\normalsize
\noindent \textbf{Observation of a second junction-TLS.} We observed a further example of qubit readout failures due to junction-TLS in the investigated sample and strain range. In Fig.~9a, we plot the raw data as shown in Fig.~1c in a wider range and at higher contrast. Besides the readout failure at a piezo voltage of $V_p$ = 55\,V, another one can be seen at $V_p$ = 69\,V. Closer inspection of the data in Fig.~9a and the extracted fit positions shown in Fig.~9b reveals a faint TLS trace, whose linear extrapolation (blue line) matches the resonator frequency when the readout failure is observed. This second TLS caused spoiled qubit readout in a broader strain range but did not display a pronounced anti-crossing with the qubit, which may be explained by a higher decoherence rates of TLS 2.\\[2cm]

\onecolumngrid
\phantom{greetings from Ferdi}\\
\includegraphics[width=\textwidth]{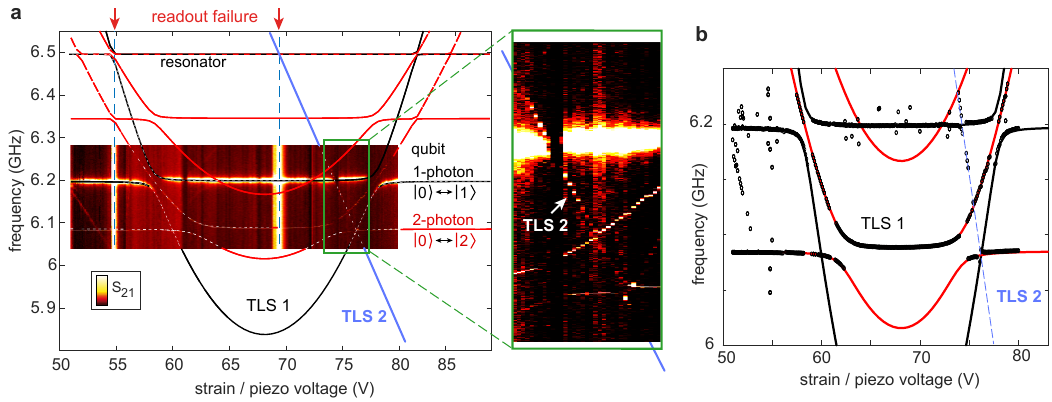}
\small{Fig. 9 a) Raw data of Fig.~1c, plotted at higher contrast. Black and red lines are the fits to 1- and 2-photon transitions of TLS 1 as discussed in the main text. The blue line is a linear fit to the faintly visible trace of TLS 2. Both TLS cause readout failure (red arrows).
	b) Detected resonance frequencies (points) and QuTiP simulation for TLS 1 with best-fitting parameters (solid lines).}	\\
\twocolumngrid
\phantom{phantom}
\section*{Bibliography} \vspace{-0.2cm}
\bibliographystyle{unsrt}
\bibliography{Biblio}

\end{document}